\documentclass[reprint,amsmath,amssymb,prl,superscriptaddress]{revtex4-1}

\usepackage{graphicx}
\usepackage{dcolumn}
\usepackage{bm}
\usepackage{hyperref,xr}
\usepackage{color}

\begin{document}

\title{Model of collective fish behavior with hydrodynamic interactions}
\author{Audrey Filella}
\affiliation{Aix Marseille Univ, CNRS, Centrale Marseille, IRPHE, 13013 Marseille, France}
\author{Fran\c cois Nadal}
\affiliation{Department of Mechanical, Electrical and Manufacturing Engineering, Loughborough University, Loughborough LE11 3TU, UK}%
\author{Cl\'ement Sire}
\affiliation{CNRS, Universit\'e Paul Sabatier, Laboratoire de Physique Th\'eorique, 31062 Toulouse, France}%
\author{Eva Kanso}
\affiliation{Aerospace and Mechanical Engineering, University of Southern California, 854 Downey Way, Los Angeles, CA 90089, USA}%
\author{Christophe Eloy}%
\email{eloy@irphe.univ-mrs.fr}
\affiliation{Aix Marseille Univ, CNRS, Centrale Marseille, IRPHE, 13013 Marseille, France}

\date{\today}

\begin{abstract} 
Fish schooling is often modeled with self-propelled particles subject to phenomenological behavioral rules. 
Although fish are known to sense and exploit flow features, these models usually neglect hydrodynamics. 
Here, we propose a novel model that couples behavioral rules with far-field hydrodynamic interactions. 
We show that 
(1) a new ``collective turning'' phase emerges;
(2) on average individuals swim faster thanks to the fluid;
(3) the flow enhances behavioral noise.
The results of this model suggest that hydrodynamic effects should be considered  to fully understand the collective dynamics of fish.
\end{abstract} 

\maketitle

Collective animal motion is ubiquitous: insects swarm, birds flock, hoofed vertebrates herd, and even humans exhibit coordination in crowds  \cite{Reynolds1987,Couzin2003,vicsek2012,Moussaid2011,Moussaid2012}. Among these fascinating collective behaviors, schooling refers to the coordinated motion of fish. It is exhibited by half of the known fish species during some phase of their life cycle \cite{Shaw1978} and can generate different disordered phases, such as swarming (important cohesion but low polarization of fish heading), or ordered phases: milling (torus or vortex pattern), bait ball (dense ``ball'' of fish), or highly polarized schools \cite{Lopez2012}.

Interestingly, these collective phases can be achieved without any leader in the group. This has first been observed in experiments \cite{Radakov1973}, and later confirmed by the development of self-propelled particle (SPP) models \cite{vicsek1995,Gregoire2003}. In the context of fish schooling, these mathematical models have generally been constructed by assuming simple phenomenological behavioral rules, such as the popular ``three-A rules'' of avoidance, alignment, and attraction \cite{Aoki1982,Huth1992,Couzin2002}. From a physical point of view,
these SPP models have an obvious interest because of their simplicity and universality \cite{Odor2004}, and because they allow the derivation of continuum equations \cite{Toner2005}.
Similar approaches have been used in soft active matter (e.g., bacteria swarms and microtubule bundles) to derive continuous models taking into account hydrodynamic interactions at vanishing Reynolds number \cite{Marchetti2013,Saintillan2015}.
However, they have rarely been connected quantitatively to experimental observations \cite{Huth1994}.
It is only recently that it has been possible to infer and model the actual behavioral rules from the individual tracking of fish in a tank \cite{gautrais2012,calovi2014,calovi2017}.

Schooling likely serves multiple purposes \cite{Krause2002}, including better foraging for patchy resources and increased protection against predators.
Fish are also thought to benefit from the hydrodynamic interactions with their neighbors \cite{Weihs1973a,Alben2012,Tsang2013,Hemelrijk2015}, but it is unclear whether this requires particular configurations or regulations.
When swimming in a structured flow, fish can exploit near-field vortices generated by other fish to reduce the energetic costs of locomotion \cite{Liao2003,Liao2003a,Beal2006}.  Fish use their lateral line, a hair-based sensor running along their side \cite{Bleckmann1994}, to sense the surrounding flow, which has also proved crucial for collective behavior \cite{Partridge1980,Faucher2010}.
Yet, the existing behavioral models of fish schooling do not include hydrodynamics.
Here, we propose to combine a data-driven attraction-alignment model \cite{gautrais2012,calovi2014,calovi2017}, with far-field hydrodynamic interactions.
In the context of this new model, several questions arise:
Can the swimmers exploit the flow to swim faster on average?
Do the hydrodynamic interactions give rise to novel collective phases?
Does the flow play the role of a self-induced noise, as it is the case at low Reynolds number \cite{Ryan2011}?


%
\begin{figure}[b]
\includegraphics[width=.48\textwidth]{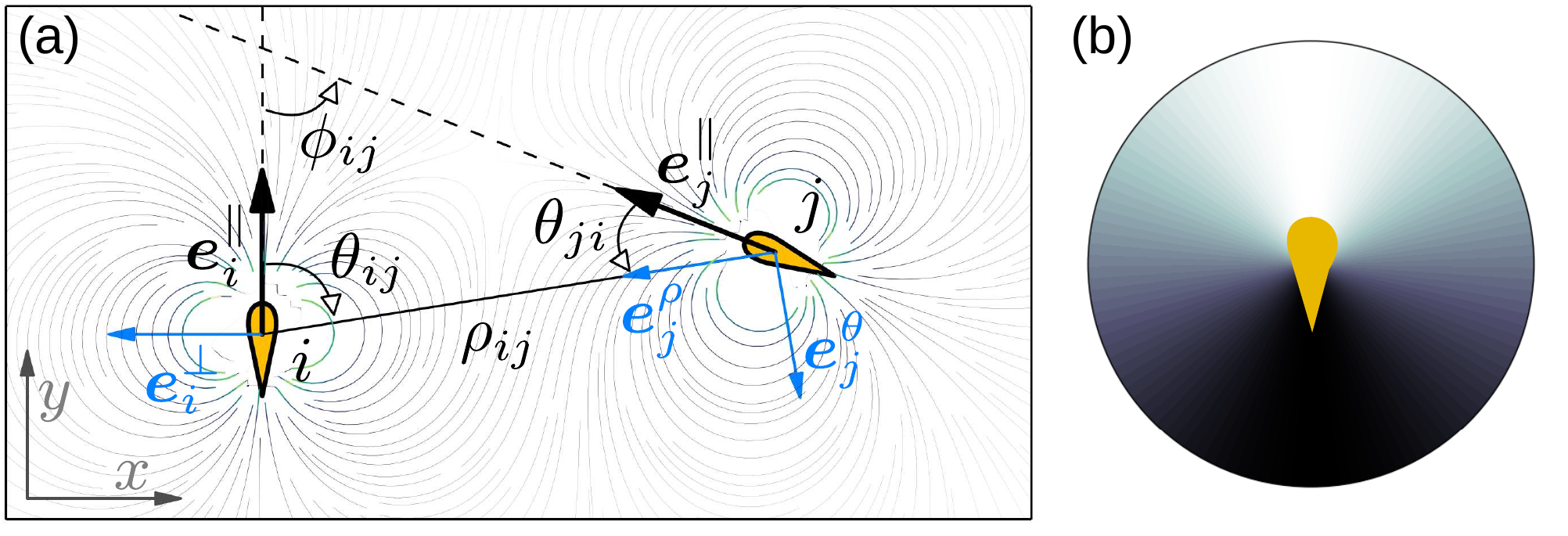} 
\caption{(a) Sketch of two interacting swimmers, showing the heading $\bm{e}^{\parallel}_i$ of swimmer $i$, its viewing angle  $\theta_{ij}$, the relative alignment angle $\phi_{ij}$, the inter-swimmer distance $\rho_{ij}$, and the polar coordinates in the reference frame of swimmer $j$ $(\bm{e}_j^{\rho}$,$\bm{e}_j^{\theta})$. (b) Grayscale representation of the anisotropic visual perception, modeled by the term $\left(1+\cos \theta_{ij}\right)$ in Eqs.~(\ref{eq:theta}--\ref{eq:average_Voronoi}).}
\label{fig_sketch}
\end{figure}

Fish are modeled as self-propelled particles moving in an unbounded two-dimensional plane. They move at constant speed $v$ relative to the flow and exhibit no inertia \footnote{Inertia is assumed to be negligible, as it was observed in the experiments made on \textit{Kuhlia mugil} \cite{gautrais2012} ($\alpha \ll 1$ in Ref. \cite{calovi2014}).}. 
Following behavioral rules inferred from shallow-water tracking experiments  \cite{gautrais2012,calovi2014}, we consider that each individual is attracted to its Voronoi neighbors with intensity $k_p$ (units m$^{-1}\,$s$^{-1}$), tends to align with the same neighbors with intensity $k_v$ (units m$^{-1}$), and is subject to a rotational noise with standard deviation $\sigma$ (units rad$\,$s$^{-1/2}$). 
Moreover, each swimmer responds to the far-field flow disturbance created by all other swimmers. This flow is an elementary dipole, with dipole intensity $Sv$ in two dimensions, where $S=\pi r_0^2$ is the swimmer surface and $r_0$ its typical length \cite{Lighthill1986,lighthill1991,Tchieu2012,Tsang2013}. 
Note that, except for hydrodynamic interactions involving the swimmer size $r_0$, swimmers are considered as point-like particles.

We use $v$ and $k_p$ to make the problem dimensionless, yielding the length scale $\sqrt{v/k_p}$ and time scale $1/\sqrt{vk_p}$. To this end, $I_{\parallel} = k_v \sqrt{v/k_p}$, $I_n = \sigma (vk_p)^{-1/4}$ and $I_f = S k_p / v$ characterize the alignment, noise and dipole intensities, respectively. 
The  dimensionless equations of motion are
\begin{eqnarray}
\dot{\bm{r}}_i & =  & \bm{e}^{\parallel}_i + \bm{U}_i, \label{eq:r} \\[1ex]
\dot{\theta}_i & =  &  \langle \rho_{ij} \sin(\theta_{ij}) + I_{\parallel}\sin(\phi_{ij})\rangle  + I_n \eta + \Omega_i. \label{eq:theta}
\end{eqnarray}
Equation~\eqref{eq:r} expresses that each individual, located at $\bm{r}_i$, is moving with a constant unit speed along its orientation $\bm{e}^{\parallel}_i$ (Fig.~\ref{fig_sketch}a). An additional drift term, $\bm{U}_i$, arises when hydrodynamic interactions are taken into account. 
A far-field approximation is used to model the flow \cite{Tchieu2012,gazzola2016}. Here, we choose to neglect the vorticity shed in the swimmer wakes \cite{Lauder2002,Becker2015} to keep the model simple and tractable. 
Under this potential flow approximation, each swimmer generates a dipolar flow field, and we can use the principle of  superposition to calculate the flow $\bm{U}_i$ experienced by a swimmer
\begin{equation}
\bm{U}_i =  \sum_{j\neq i} \bm{u}_{ji},~~ \mbox{with }\bm{u}_{ji}= \frac{I_f}{\pi }  \frac{\bm{e}^{\theta}_j \sin\theta_{ji}  + \bm{e}^{\rho}_j \cos\theta_{ji}}{\rho_{ij}^2},
\label{eq6}
\end{equation}
where $\bm{u}_{ji}$ is the velocity induced by swimmer $j$ at the position $\bm{r}_i$ and $(\bm{e}_j^{\rho}$,$\bm{e}_j^{\theta})$ are the polar coordinates in the  framework of swimmer $j$ (Fig.~\ref{fig_sketch}a).
The angular velocity in  Eq.~\eqref{eq:theta}  is the sum of an attraction term, an alignment term, a standard Wiener process $\eta(t)$, describing the spontaneous motion of the fish and modeling its ``free will'', and a rotational term $\Omega_i$ induced by hydrodynamic interactions.
The behavioral terms (attraction and alignment) are averaged over the Voronoi neighbors, noted $\mathcal{V}_i$, with the weight $\left(1+\cos \theta_{ij}\right)$ modeling continuously a rear blind angle \cite{calovi2014} (Fig.~\ref{fig_sketch}b)
\begin{equation}
\langle \circ \rangle  =  {\sum_{j\in \mathcal{V}_i} \circ\, (1+\cos \theta_{ij}) } \bigg/ {\sum_{j\in \mathcal{V}_i} (1+\cos \theta_{ij})}. \label{eq:average_Voronoi}
\end{equation}
The rotation induced by hydrodynamic interactions is not
due to vorticity, since it is zero for a potential flow, but to gradients of normal velocities along the swimming direction. In other words, the angular velocity due to the flow is
\begin{equation}
\Omega_{i} = \sum_{j\neq i} \bm{e}^{\parallel}_i \cdot \bm{\nabla u}_{ji} \cdot \bm{e}^{\perp}_i. \label{eq:omega}
\end{equation}

\begin{figure}[t]
\includegraphics[width=0.48\textwidth]{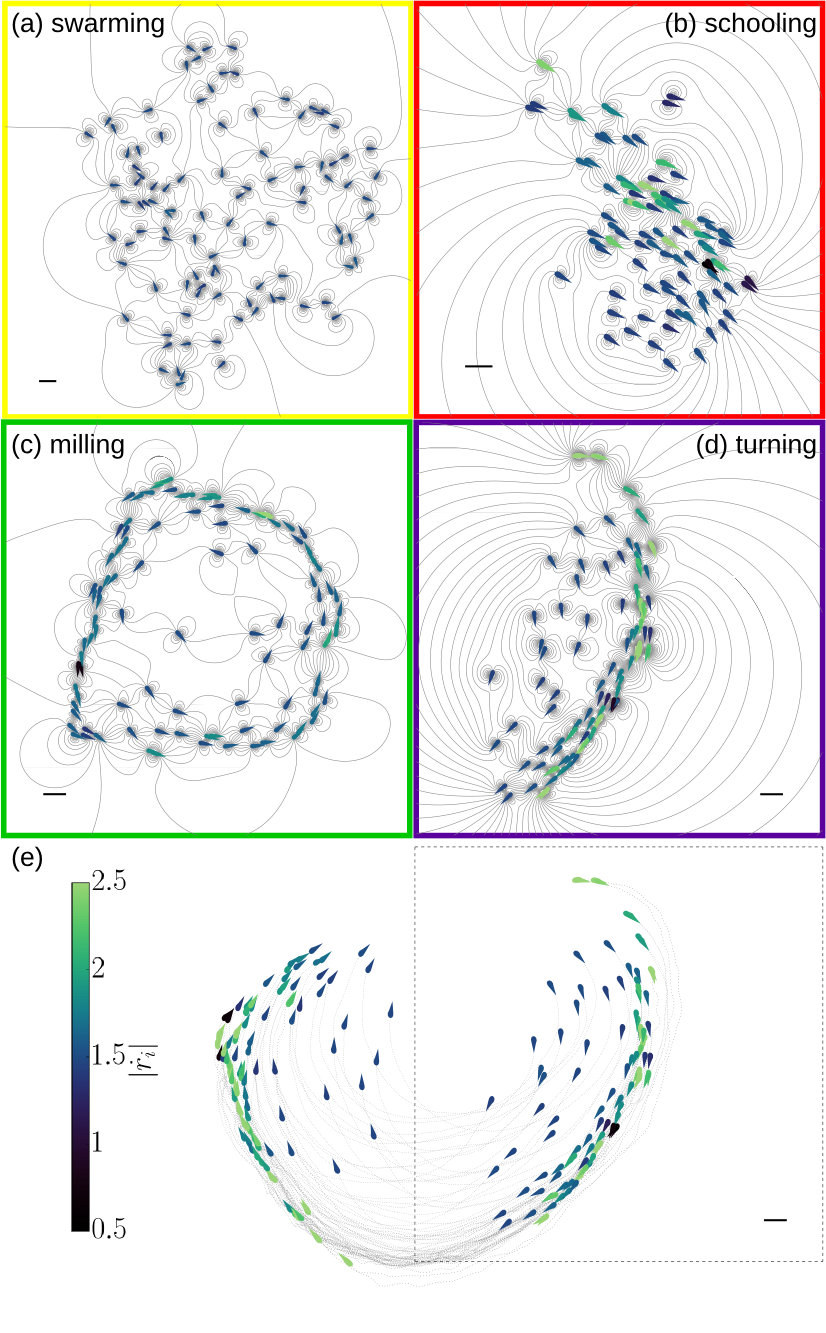} 
\caption{
(a--d) Plots of the swimmer positions and associated streamlines for different values of the parameters. On all figures, the dipole intensity is $I_f=10^{-2}$, the scale bar corresponds to $10\, r_0$, and color scale represents the instantaneous velocity.
For clarity, swimmers are represented as ``airfoils'' of length $7r_0$. 
Four distinct dynamical phases are observed:
(a) swarming for $I_n = 0.8$, $I_\parallel = 0.5$;
(b) schooling for $I_n = 0.5$, $I_\parallel = 9$;
(c) milling for $I_n = 0.3$, $I_\parallel = 1.5$; and
(d) turning for $I_n = 0.2$, $I_\parallel = 4$.
(e) Paths followed by each swimmer during 12 dimensionless time units, the final time corresponding to (d).  
For long-time dynamics, see Supplementary Fig.~1 and Supplementary Movies 1--4.}\label{fig:schooling_phases}
\end{figure}
\begin{figure*}[!thb]
\includegraphics[width=1\textwidth]{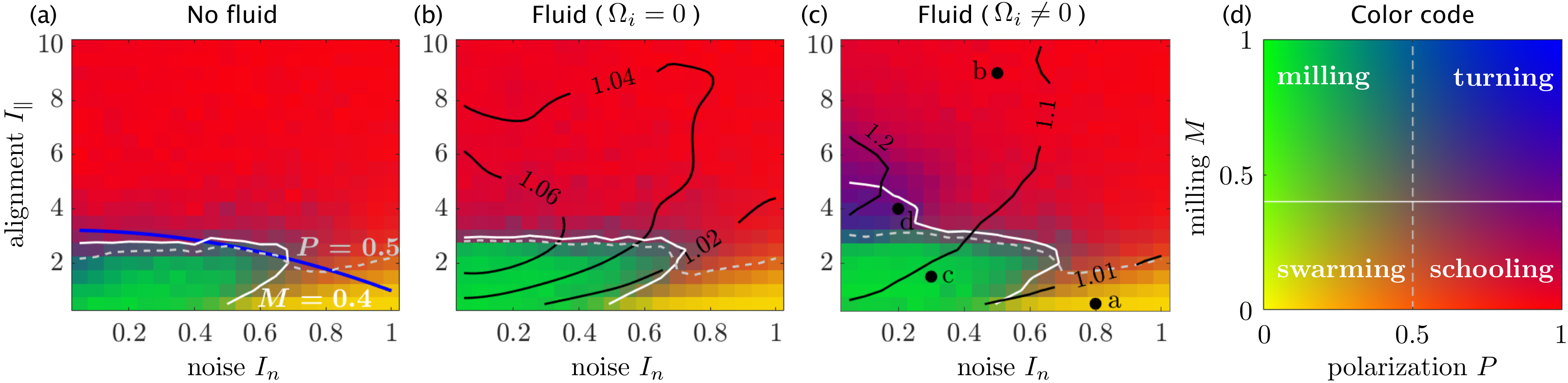} 
\caption{Phase diagram using the value of the polarization $P$ and the milling $M$ as a color code (d), for three cases: (a) no fluid ($\bm{U}_i=\Omega_i=0$); (b) fluid without induced rotation  ($\bm{U}_i\neq0$, $\Omega_i=0$); (c) full hydrodynamic model ($\bm{U}_i\neq0$, $\Omega_i\neq0$).
In all cases, the dipole intensity is $I_f=10^{-2}$, the solid white line indicates the $M=0.4$ level, and the dashed line the $P=0.5$ level. Solid black lines in (b--c) show the $V$-levels  and the solid blue line in (a) shows the milling-schooling transition line found in \cite{calovi2014}.
In (c), the black dots show the parameter values corresponding to Fig.~\ref{fig:schooling_phases}a--d.}
\label{fig:phase_space}
\end{figure*}

Another interpretation of this angular velocity is to consider that each swimmer is a dumbbell oriented in the swimming direction, with each weight advected by the flow (Supplementary Fig.~2).
The intensity of the hydrodynamic interactions (related to both the drift term $\bm{U}_i$ and the induced rotation $\Omega_{i}$) is proportional to the dipole intensity $I_f$. For the $10\,$cm-long fish ($r_0=5\,$cm) considered in Ref. \cite{gautrais2012}, the cruise speed is $v \approx 0.2\,$m$\,$s$^{-1}$, and $k_p=0.41\,$m$^{-1}\,$s$^{-1}$, yielding a dipole intensity $I_f \approx 0.016$.
In the present model, the flow induces translational and rotational motions, whose origin are \emph{physical}, but, in principle, it could also elicit a \emph{behavioral response} (\emph{e.g.}, a tendency for fish to go along or against the flow) \cite{calovi2017}.



We consider a group of $N=100$ individuals, with random initial orientations and initially distributed in a $20\times 20$ box (in dimensionless length units), although the subsequent dynamics is not affected by the initial conditions.
The dynamical system described by Eqs.~(\ref{eq:r}--\ref{eq:omega}) is solved numerically, using an explicit scheme with time step $\delta t = 10^{-2}$.
Depending on the values of the three dimensionless parameters ($I_\parallel$, $I_n$, and $I_f$), four different dynamical phases emerge (Fig.~\ref{fig:schooling_phases}).
When noise is comparable or larger than the alignment, we observe a disordered \emph{swarming} phase (Fig.~\ref{fig:schooling_phases}a): swimmers form a sparse group with no preferential orientation.
When alignment intensity is stronger, the group is denser and individuals tend to swim in the same direction: this is known as the \emph{schooling} phase (Fig.~\ref{fig:schooling_phases}b).
When alignment and attraction are comparable and noise is low or moderate, the group reaches a \emph{milling} phase (Fig.~\ref{fig:schooling_phases}c): it forms a  ``vortex''.
These three phases (swarming, schooling, and milling) can also be observed without any hydrodynamic interactions \cite{calovi2014} ($\bm{U}_i=\bm{0}$ and $\Omega_i=0$, in Eqs.~(\ref{eq:r}--\ref{eq:theta})).
However, when the flow is explicitly taken into account, a novel phase appears that we call the \emph{turning} phase (Fig.~\ref{fig:schooling_phases}d--e). In this new phase, swimmers tend to align along a preferential orientation and, at the same time, the group follows a large-scale quasi circular trajectory. 


In order to precisely characterize these different phases, the global order parameters $P$ and $M$ are introduced \footnote{Following the usage in collective behavior studies, we refer to the different spatial organizations of the swimming group as ``phases'', which should not to be confused with the phases used in equilibrium statistical physics. Here, they correspond to different regions in the space of order parameters ($P$, $M$), defined with some arbitrariness when the system is finite.}, along with the average speed $V$
\begin{equation}
P = | \overline{ \bm{e}^{\parallel}_i } |,\quad
M = \frac{\left|\overline{ \bm{e}^r_i\times\bm{\dot{r}}_i} \right|} { \left|\overline{\bm{e}^r_i}\right| \left|\overline{\bm{\dot{r}}_i} \right| },\quad
V = \overline{\left|\bm{\dot{r}}_i \right|} ,
\end{equation}
where $\bm{e}^r_i = (\bm{r}_i - \overline{\bm{r}_i}) / |\bm{r}_i - \overline{\bm{r}_i}|$ is the unit vector along the segment joining the center of mass of the group and the $i$-th swimmer, and the over bar denotes average over all individuals.
The parameter $P$ is the polarization, $M$ is
the milling and corresponds to the normalized angular momentum of the group (straight-line schooling gives a value of 0, while perfect milling gives 1).

To assess the importance of hydrodynamic interactions, we performed a systematic parametric study for three cases (Fig.~\ref{fig:phase_space}): a pure behavioral model with no effects of the fluid ($\bm{U}_i=\bm{0}$ and $\Omega_i=0$, in Eqs.~(\ref{eq:r}--\ref{eq:theta})), a simple model of hydrodynamic drift without induced rotation ($\bm{U}_i\ne\bm{0}$ and $\Omega_i=0$), and a full hydrodynamic model with both induced translation and rotation ($\bm{U}_i\ne\bm{0}$ and $\Omega_i\ne 0$). For each set of parameters, $P$, $M$, and $V$ are obtained after time-averaging over $\Delta t=100$ (after waiting 100 time units to ensure that the transient dynamics of few dimensionless time units is over), and ensemble-averaging over 100 realizations.

In the absence of hydrodynamic interactions, the results of Ref. \cite{calovi2014} are recovered (Fig.~\ref{fig:phase_space}a). 
For $P>0.5$, which roughly correspond to $I_\parallel\gtrsim 2$, we observe the schooling phase. For $M>0.4$, obtained for
$I_\parallel\lesssim 2$ and $I_n\lesssim 0.5$, the group exhibits a milling phase. 
For all other cases tested, the swarming phase is observed. 
We chose the threshold values $P=0.5$ and $M=0.4$ to distinguish the four phases, but $P$ and $M$ vary continuously in the parameter space. An alternative choice of thresholds would be possible and would yield qualitatively similar results.

When hydrodynamic drift is introduced with $I_f=10^{-2}$, but induced rotation is neglected (Fig.~\ref{fig:phase_space}b), the phase diagram is practically unchanged. The only difference is that the mean velocity $V$ is now slightly greater than 1 ($V=1$ when hydrodynamic interactions are neglected).
When the full hydrodynamic model is considered (Fig.~\ref{fig:phase_space}c), the new turning phase appears for $M>0.4$ and $P>0.5$ (corresponding to $I_n\lesssim0.25$ and $3\lesssim I_\parallel\lesssim 5$) and $V$ is increased. Looking at the swimming speeds of each individual in the group (Fig.~\ref{fig:schooling_phases}e), we see that some individuals swim slower because of the fluid, and some others swim much faster with swimming speed reaching $\left|\bm{\dot{r}}_i \right| = 2.5$.



\begin{figure}[!tb]
\includegraphics[width=0.48\textwidth]{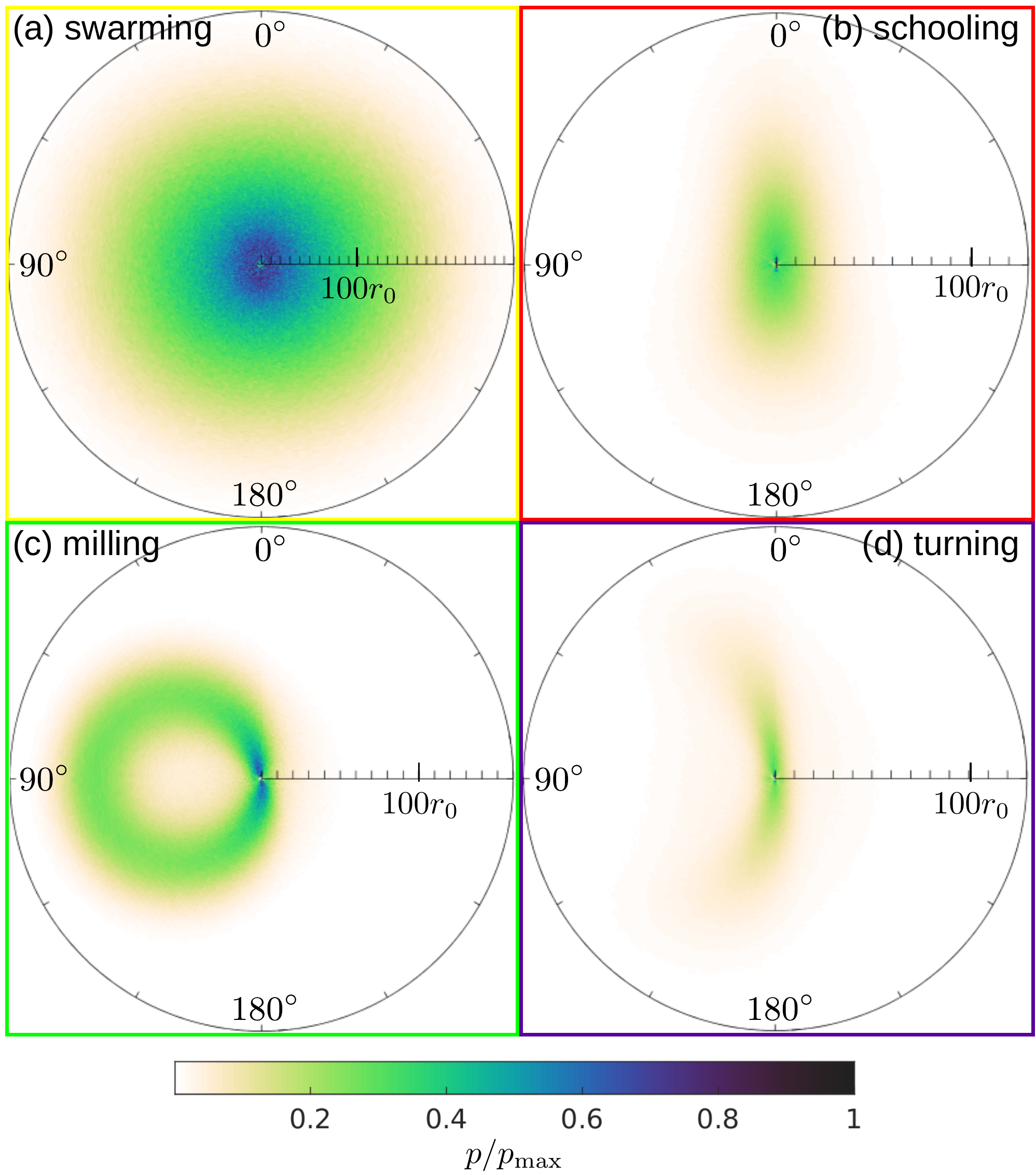} 
\caption{Heat maps showing the probability of presence $p(\rho,\theta)$ of all other swimmers in the framework of an individual. The parameters are the same as in Fig.~\ref{fig:schooling_phases}.}
\label{fig:heatmap}
\end{figure}

To understand why individuals always swim faster on average when hydrodynamic interactions are taken into account, we computed the probability of presence of other swimmers in the framework of each individual (Fig.~\ref{fig:heatmap}).
For the same parameter values as in Fig.~\ref{fig:schooling_phases}, we  collected the position and orientation of each swimmer during $\Delta t =900$.
For the swarming phase (Fig.~\ref{fig:heatmap}a), the probability of presence is isotropic. Hence, there is no velocity increase due to dipolar hydrodynamic interactions on average.
However, for the schooling phase, individuals tend to swim in-line rather than side-by-side, leading to a density distribution  polarized along the vertical (Fig.~\ref{fig:heatmap}b), or equivalently $\theta_{ji}$ preferentially around $0^\circ$ or $180^\circ$ (Fig.~\ref{fig_sketch}). This induces a velocity increase along the swimming direction $\bm{e}^{\parallel}_i$ (see Eq.~\eqref{eq6}). The same is true for the milling and turning phases (Fig.~\ref{fig:heatmap}c--d).

Why do individuals tend to swim in-line in the presence of the fluid? To address this question, we examined the preferential location of the nearest Voronoi neighbors in the swarming and schooling phases, in the presence of the fluid or not (Supplementary Fig.~3).
It appears that the preferential in-line configuration is only present in the full hydrodynamic model. This is because the side-by-side configuration becomes unstable when hydrodynamic interactions are considered \cite{Kanso2014,Becker2015} (Supplementary Fig.~4), and swimmers thus tend to spend more time in-line.

\begin{figure}[b]
\includegraphics[width=0.48\textwidth]{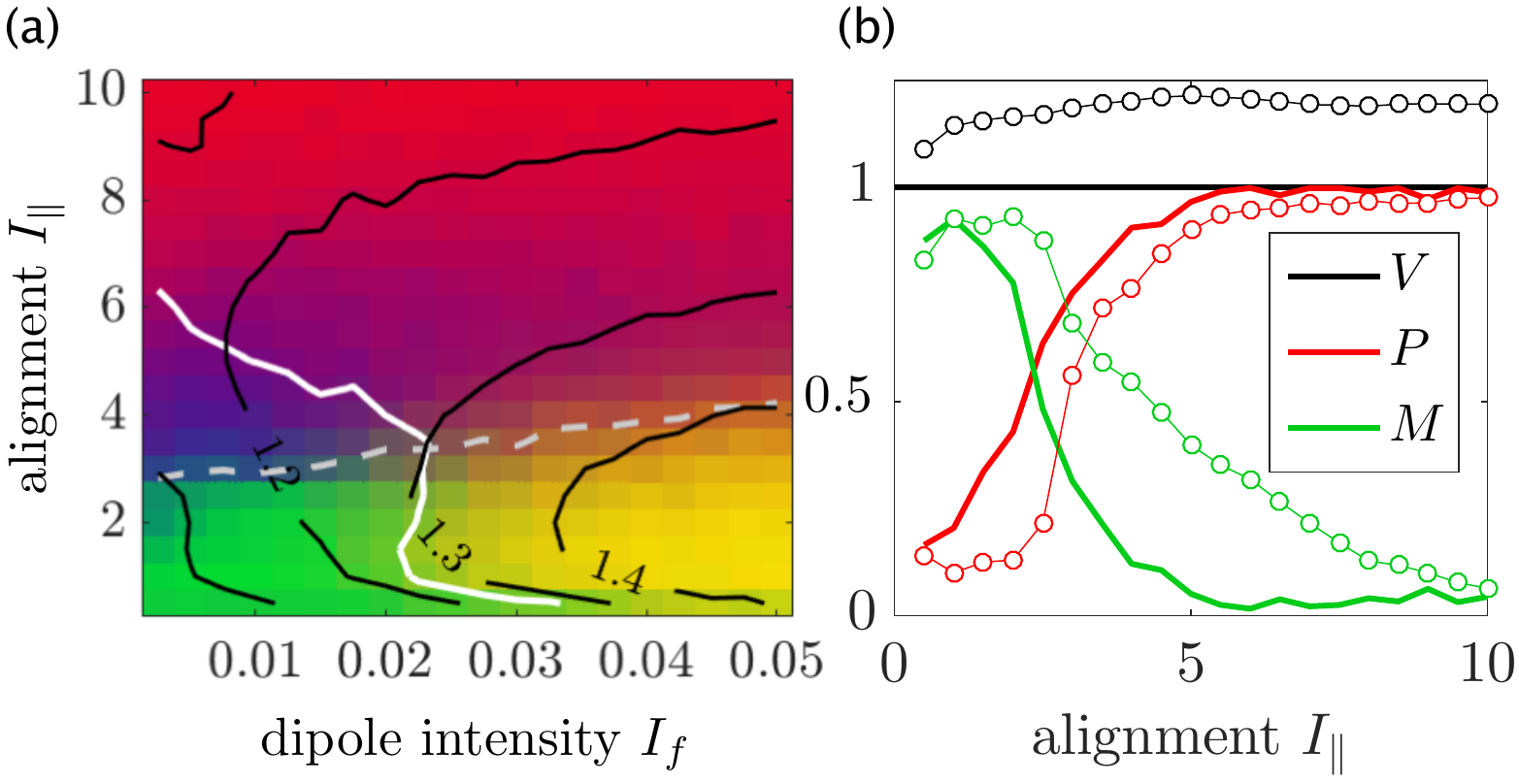} 
\caption{(a) Phase diagram in the absence of noise ($I_n=0$). The color code and the contour levels are the same as in Fig.~\ref{fig:phase_space}.
(b) Values of the polarization $P$ (red), milling $M$ (green), and mean velocity $V$ (black) for two cases: no fluid ($I_f=0$) and low noise ($I_n=0.05$) with solid lines; full hydrodynamic model ($I_f=10^{-2}$) and no noise ($I_n=0$) with open symbols. }
\label{fig:no_noise}
\end{figure}

The role of the fluid is not only to increase the swimming speed on average, but also to introduce a source of disorder. To assess if this disorder has the same effective impact as the noise $I_n\eta$ in Eq.~\eqref{eq:theta} that describes the spontaneous angle fluctuations, we performed simulations with no noise ($I_n=0$) and with varying dipole intensity $I_f$ (Fig.~\ref{fig:no_noise}a).
The phase diagrams shown in Fig.~\ref{fig:phase_space}c and Fig.~\ref{fig:no_noise}a are qualitatively similar, both exhibiting the four phases (schooling, swarming, milling, and turning) with the same topology.
It thus shows that the hydrodynamic interactions also play the role of a rotational noise.
There are however some differences between Fig.~\ref{fig:phase_space}c and Fig.~\ref{fig:no_noise}a. 
First, the average velocity increases when the dipole intensity increases whereas it tends to decrease with noise intensity. 
Second, for large $I_f$ and small $I_\parallel$, even if $P$ and $M$, are both small, the school does not behave as in the swarming phase. It can be composed of very dense clusters of quasi static swimmers (Supplementary Fig.~5). This non-realistic behavior is an artifact of the simulations due to the absence of noise.

There is a continuous transition between the milling, turning, and schooling phases (Figs.~\ref{fig:phase_space} and \ref{fig:no_noise}). 
In a pure behavioral model (with no fluid), the transition between the milling and schooling is also continuous, but $M$ is systematically higher when the fluid is present (Fig.~\ref{fig:no_noise}b).
Although the turning and milling phases are similar, their origins are different.
As Calovi \emph{et al.} \cite{calovi2014} already noted, the milling phase can only be stabilized when swimmers have an anisotropic visual perception (Fig.~\ref{fig_sketch}b and Eq.~\eqref{eq:average_Voronoi}).
On the contrary, the turning phase still exists when visual perception is isotropic (Supplementary Fig.~7), but requires the full hydrodynamic model to be observed.
Although experimental data are too scarce to support the existence of the turning phase in real fish schools, we can speculate that the speed enhancement achieved in this phase could be advantageous to the group, for energetic considerations, or  when confronted to a danger.
Note that the milling and the turning phases, when the number of swimmers is large, can break into several smaller groups and thus affect the value of the order parameters $P$ and $M$ (Supplementary Movies 5-7, and Supplementary Fig.~6).


In summary, we proposed a new model of collective fish motion that includes behavioral rules and far-field hydrodynamic interactions.
By simulating numerically the dynamics of this model for a large group of swimmers, we showed that, on average, fish swim faster in a school, due to the presence of the fluid. This suggests that fish would need less energy to swim in a school for a given swimming speed. 
This emergent property results from the preferential in-line pairing of swimmers, more robust than the side-by-side configuration.
In addition, we observed a new phase, called the turning phase, which only exists with the full hydrodynamic model.
Finally, we showed that the fluid has similar effect to the spontaneous cognitive rotational noise.
These promising results underline the importance of hydrodynamic interactions in fish schooling.
In future work, it will be important to assess the validity of the far-field approximation used here by integrating the fish wakes into the fluid model.

\begin{acknowledgments}
We are grateful to G.~Theraulaz for his valuable insight. A.\,F. 
acknowledges support from A*MIDEX (ANR-11-IDEX-0001-02), and the Labex MEC (ANR-10-LABX-0092).
E.\,K. acknowledges sabbatical support from the Flatiron Institute at the Simons Foundation and research support from Office of Naval Research (ONR) through grants N00014-14-1-0421 and N00014-17-1-2287 and the Army Research Office (ARO) through the grant W911NF-16-1-0074.
\end{acknowledgments}


%

\end{document}